%% file: mainpaper.tex
\documentclass[runningheads]{llncs}

\usepackage{graphicx}
\usepackage{epsfig}
\usepackage{amsmath}
\usepackage{amssymb}
\usepackage[dvipsnames]{xcolor}
\usepackage{enumerate} 
\usepackage{amssymb}
\usepackage{pifont} 
\usepackage{tabularx}
\usepackage{relsize}

\usepackage{array}
\usepackage{booktabs}
\usepackage[normalem]{ulem}
\usepackage{tabularray}
\usepackage{colortbl}
\usepackage{tablefootnote}
\usepackage[flushleft]{threeparttable}

\usepackage[pdfstartview=XYZ,
bookmarks=true,
colorlinks=true,
linkcolor=black,
urlcolor=black,
citecolor=green,
pdftex,
bookmarks=true,
linktocpage=true, 
hyperindex=true
]{hyperref}

\usepackage{orcidlink}

\newcommand{\xxmark}{\ding{53}}%

\newcolumntype{h}{>{\hsize=1.1\hsize}X}
\newcolumntype{b}{>{\hsize=.45\hsize}X}
\newcolumntype{s}{>{\hsize=.14\hsize}X}

\makeatletter
\def\blfootnote{\xdef\@thefnmark{}\@footnotetext}
\makeatother

\begin{document}

\title{CheXtriev: Anatomy-Centered Representation for Case-Based Retrieval of Chest Radiographs}
\blfootnote{N. Akash and A. Tadanki contributed equally.}

\titlerunning{CheXtriev: Anatomy-Centered Case-Based Retrieval of CXRs}

\author{Naren Akash R J \index{Akash, Naren}\orcidlink{0000-0002-5218-8434}\and
Arihanth Tadanki \index{Tadanki, Arihanth}\orcidlink{0009-0006-5554-1703} \and
Jayanthi Sivaswamy \index{Sivaswamy, Jayanthi}\orcidlink{0000-0001-9588-0482}
}

\authorrunning{N. Akash et al.}

\institute{Center for Visual Information Technology, \\International Institute of Information Technology Hyderabad, India
\email{naren.akash@research.iiit.ac.in}\\
\href{https://cvit.iiit.ac.in/mip/projects/chextriev}{\email{https://cvit.iiit.ac.in/mip/projects/chextriev}}
}

\maketitle              

\input{sections/00_abstract}
\input{sections/01_introduction}

\input{sections/02_methods}

\input{sections/03_experiments}

\input{sections/04_discussion}

\input{sections/05_conclusion}
\input{sections/06_bibliography}

\input{SupplementaryFile-3636/SupplementaryLatexSource-3636/supplementarypaper}

\end{document}

%% file: sections/00_abstract.tex
\begin{abstract}
We present CheXtriev, a graph-based, anatomy-aware framework for chest radiograph retrieval. Unlike prior methods focussed on global features, our method leverages graph transformers to extract informative features from specific anatomical regions. Furthermore, it captures spatial context and the interplay between anatomical location and findings. This contextualization, grounded in evidence-based anatomy, results in a richer anatomy-aware representation and leads to more accurate, effective and efficient retrieval, particularly for less prevalent findings. CheXtriv outperforms state-of-the-art global and local approaches by $18\%$ to $26\%$ in retrieval accuracy and $11\%$ to $23\%$ in ranking quality. The code is available at \url{https://github.com/cvit-mip/chextriev}.
\end{abstract}

%% file: sections/01_introduction.tex
\section{Introduction}


In clinical practice, expertise accumulates with experiences \cite{experience_aaai1982}. Clinicians often use case-based reasoning \cite{cbr_ai1991} to understand and diagnose patients by drawing parallels between past cases and current presentations. This analogy-driven approach hinges on effectively retrieving relevant past cases, making medical image retrieval (MIR) a cornerstone of evidence-based clinical decision-making. MIR facilitates the identification of similar past cases, allowing clinicians to: (i) refine diagnosis by suggesting investigative features or proposing alternative diagnoses, (ii) optimize treatment planning based on past outcomes, and (iii) enhance explanation and failure recovery by identifying clinically relevant past errors and their corrective actions \cite{star_tsmc1987}. As such, robust MIR systems hold significant potential to improve patient outcomes and advance medical care.\\


\noindent
Despite its promise, MIR presents distinct challenges \cite{survey_media2018}. Unlike general-purpose image retrieval, which focuses on global image regions, MIR has to contend with images that exhibit similar global features across patients, with subtle, fine-grained markers serving as critical disease indicators. This is particularly evident in chest radiographs, the most common yet challenging imaging modality. Its interpretation is prone to errors due to the two-dimensional projection of three-dimensional structures, leading to the superimposition of anatomies, low-contrast features, and multiple subtle, non-specific abnormalities. Studies suggest error rates as high as 30\% in patients with 
abnormal findings 
and with up to 56\% disagreement among radiologists \cite{cmf_chest2023}. MIR, especially for cases with multiple findings, remains understudied despite major strides in computer vision for multi-label settings \cite{multilabelsurvey_air2020}. \\


\noindent
Automated case-based retrieval of chest radiographs necessitates learning \textit{discriminative yet informative representations} that can help bridge the gap between low-level visual features and high-level clinical interpretations. Hashing, widely used for computational and storage efficiency, maps high-dimensional features to compact hash codes. Advances in deep learning offer end-to-end training of deep hashing models such as DRH \cite{drh_miccai2017}, OSDH \cite{osdh_miccai2018} and SH-EBM \cite{shebm_miccai2021} for multi-label chest radiographs, automatically learning both features and hash codes. However, they often lose crucial information related to disease classification and the regions of interest. This has been addressed recently with attention-based triplet hashing (ATH) \cite{ath_media2021} to implicitly focus on specific regions. ATH still relies on global image representation, potentially limiting its ability to capture fine-grained clinical information. 
Other emerging cross-modal approaches, such as MMDL \cite{mmdl_miccai2021} and X-TRA \cite{xtra_ipmi2023}, incorporate text reports. Existing MIR approaches use image representations which do not fully leverage radiologists' systematic approach \cite{systematic_chest2012} to read chest radiographs, where the anatomical location of observed findings is often used for differential diagnosis. While focusing on anatomy has shown promise in classification (AnaXNet \cite{anaxnet_miccai2021}), change detection (CheXRelNet \cite{chexrelnet_miccai2022}), and report generation (RGRG \cite{report_cvpr2023}), to our best knowledge, no work has attempted to incorporate anatomical reasoning into the retrieval process itself.
\\


\noindent
In this paper, we focus on MIR for chest radiographs aiming to find relevant cases solely from visual content. Our approach addresses the limitations of current multi-label retrieval methods by leveraging anatomy-aware representation learning. Unlike existing methods that rely on global features, we explicitly target the subtle, yet critical informative details from specific anatomical regions and fuse them to learn a richer representation that incorporates the interplay between anatomical location and radiological findings. This contextualization, grounded in evidence-based anatomy\cite{eba_clincanat2014}, leads to more accurate, effective and efficient retrieval compared to methods using solely global features. Furthermore, we incorporate anatomy-aware saliency to highlight regions contributing to the retrieval decision to build model understanding and transparency.

%% file: sections/02_methods.tex
\section{Method}

\begin{figure}[t]
    \centering
    \includegraphics[width=\linewidth]{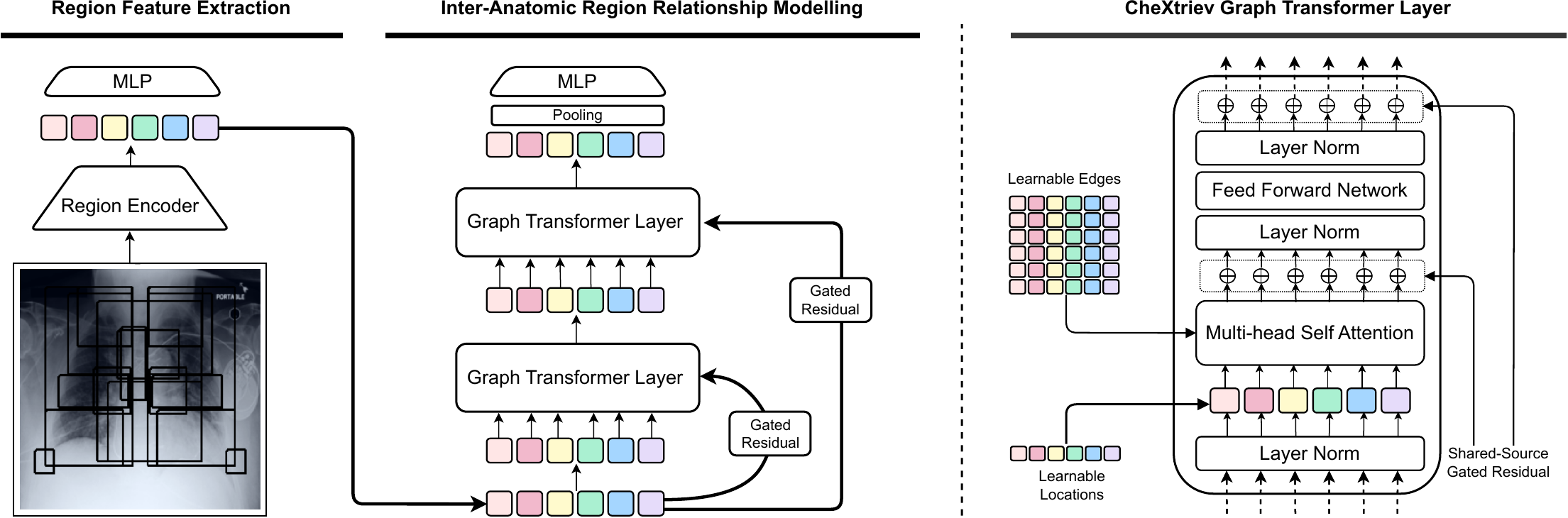}
    \caption{A schematic illustration of CheXtriev.}
    \label{fig:netowrk}
\end{figure}

Case-based chest radiograph retrieval systems identify images from a collection which are similar to a given query image. This is achieved by analyzing the visual content of the radiographs as well as focusing on the presence and location of anatomical abnormalities.
In this work, we propose a system called CheXtriev, which first extracts informative features from anatomically defined regions and constructs a graph where nodes represent regions and edges capture spatial relationships and potential co-occurrences between findings. The graph leverages learnable location and edge embeddings to capture spatial context and relationships between regions, allowing a graph transformer architecture with global edge-aware attention and gated residuals to learn robust global-local context representations for accurate, effective and efficient retrieval. An overview of our proposed pipeline is shown in Figure \ref{fig:netowrk}.

\subsection{Thoracic Radioanatomic Region Representation Extraction}
Radiologists systematically analyse chest radiographs by sequentially assessing anatomical structures and interfaces to precisely detect abnormalities. Inspired by this, we identify eighteen essential anatomical regions in frontal radiographs. These regions include the right and left lungs, encompassing their apical, upper, mid, and lower zones, as well as the right and left hilar structures, costophrenic angles, mediastinum (and its upper portion), cardiac silhouette, and trachea. We obtain the anatomical region representations $R = \{r_i\}_{i=1}^{N} \in \mathbb{R}^{N \times D_R}$, where $N$ is the number of regions 
and each ${r_i}$ of dimension $D_R$ is extracted by a fixed, pretrained ResNet50 feature extractor after the last global average pooling layer.

\subsection{Modeling Inter-Region Relationships with Graph Transformer}
The complex diagnostic reasoning process of a radiologist relies on anatomical interdependence, spatial pattern recognition, and disease co-occurrence. We design a novel graph transformer \cite{gt_aaaiws2020} framework, to capture both global and local contexts and learn latent relationships between different regions. 
\\

\noindent 
\textbf{Feature projection and location embeddings.} We project each of the region's features $r_i \in \mathbb{R}^{D_R} $ with a two-layer perceptron (MLP) with one ReLU non-linearity. We \textit{learn} $N$ location embedding vectors $E^{L} \in \mathbb{R}^{D_R \times N}$ to distinguish between one anatomical region from the other. Combining the features with corresponding embedding vectors provides a rich representation $\tilde{r}_i = \text{MLP}(r_i) + E^{L}_i$.\\

\noindent
\textbf{Edge-aware graph attention.} To model anatomical relationships, we construct a graph with \textit{learnable} continuous edges to capture inter-region dependencies. Each edge in the graph connects two regions, denoted as $j$ and $i$ (neighbour and reference region, respectively). We calculate $C$-way multi-head attention that incorporates both region features and learnable edge information as follows. \
Reference region features $h_i$ and neighbouring region features $h_j$ are transformed into query $q_{c,i}^{(l)} \in \mathbb{R}^{D_T}$ and key $k_{c,j}^{(l)} \in \mathbb{R}^{D_T}$ vectors using trainable weights $W_{c,q}^{(l)}$, $W_{c,k}^{(l)}$ and biases $b_{c,q}^{(l)}$ and $b_{c,k}^{(l)}$. 
$q_{c,i}^{(l)} = W_{c,q}^{(l)} h_i^{(l)} + b_{c,q}^{(l)}$; \ \  
$k_{c,j}^{(l)} = W_{c,k}^{(l)} h_j^{(l)} + b_{c,k}^{(l)}$. Here, 
\textit{l} corresponds to $l^{th}$ layer. Edge features $e_{ij}$ are learnt and encoded and added into the key vector as additional information for each layer, allowing the model to capture context-specific latent relationships beyond inherent node features: 
$e_{c,ij} = W_{c,e} e_{ij} + b_{c,e}$ \ and 
$\alpha_{c,ij}^{(l)} = \frac{\left\langle q_{c,i}^{(l)}, k_{c,j}^{(l)} + e_{c,ij} \right\rangle}{\mathlarger{\sum}_{u \in \mathcal{N}{(i)}} \left\langle q_{c,i}^{(l)}, k_{c,u}^{(l)} + e_{c,iu} \right\rangle}$, 
where $\langle q, k\rangle = \exp{ \left( \frac{q^T k}{\sqrt{D_T}} \right) }$ is exponential scale cosine similarity function and $D_T$ is the hidden size of each head. 
We calculate a weighted sum of neighbouring region features, incorporating both node and edge information, and make a message aggregation to the reference region as: $v_{c,j}^{(l)} = W_{c,v}^{(l)}h_j^{(l)} + b_{c,v}^{(l)}; \ \
\hat{h}_i^{(l+1)} = \Vert_{c=1}^{C} \sum_{j \in \mathcal{N}{(i)}} \alpha_{c,ij}^{(l)} ( v_{c,j}^{(l)} + e_{c,ij} )$.
\\

%

\noindent
\textbf{Multi-level feature aggregation.} We introduce shared-source gated residual connections to \textit{selectively} aggregate global-local context-aware features as: $g_i^{(l)} = W_r^{(l)}h_i^{(l)} + b_r^{(l)}$; \ 
$\beta_i^{(l)} = \text{sigmoid}( W_g^{(l))} [ \hat{h}_i^{(l)}; r_i^{(l)}; \hat{h}_i^{(l+1)} - r_i^{(l)} ])$; \ $h_i^{(l+1)} = \text{ReLU} 
 (\text{LayerNorm}( ( 1 - \beta_i^{(l)})\hat{h}_i^{(l+1)} + \beta_i^{(l+1)}h_i^{(0)})).$ 
The aggregated, hierarchical representation incorporates more comprehensive, fine-grained information at different levels of granularity to better detect abnormalities among different regions.\\

\noindent
\textbf{Classifier.} We average the multi-head output on the last layer \textit{L} of graph transformer and remove non-linearities: \ $\hat{h}_i^{(L+1)} = \frac{1}{C} \sum_{c=1}^C  \sum_{j \in \mathcal{N}{(i)}} \alpha_{c,ij}^{(L)}  
 ( v_{c,j}^{(L)} + e_{c,ij})$; \  
$h_i^{(L+1)} = (1-\beta_i^{(L)}) \hat{h}_i^{(L+1)} + \beta_i^{(L)}g_i^{(L)}$. We feed the mean-pooled features into shared linear layers for multi-label classification using class-wise binary cross-entropy loss. We perform normalization on the final layer features to obtain dense vectors that capture anatomy-aware representations for similarity search.

\subsection{Efficient Retrieval and Visual Interpretability for Similarity}

\noindent
To efficiently retrieve relevant chest radiographs, we adopt a fast and exact \textit{k}-nearest neighbour retrieval model using FAISS \cite{faiss_2024}. It builds a flat index for our encoded, dense data embeddings, enabling efficient inner product similarity search with the query vector. This non-parametric approach effectively identifies radiographs with the highest cosine similarity from the pre-built vector database.\\

\noindent
In order to visually interpret the results, we adopt an occlusion-based method \cite{xmir_wacv2022} and derive similarity-based saliency maps. Instead of using small, sliding occlusions, we directly occlude pre-defined anatomical regions within the retrieved image. We then calculate the cosine similarity between the query image and each occluded version. Regions causing significant drops in cosine similarity are deemed critical for overall similarity, while those with minimal change are considered less important. This analysis is visualized as a saliency map, highlighting the most influential regions for similarity between the query and retrieved image.


%% file: sections/03_experiments.tex
\section{Experiments}

\subsection{Data}

We use the MIMIC-CXR-JPG v2.0.0 \cite{mimiccxrjpg_2019} database for all our experiments and follow the official MIMIC-CXR \cite{mimiccxr_scidata2019} data splits. We select the frontal chest radiograph images (PA or AP view) with valid bounding box coordinates for all eighteen anatomical regions. These annotations are provided by the Chest ImaGenome v1.0.0 \cite{chestimagenome_neurips2021} dataset, which is constructed from MIMIC-CXR. This resulted in 226,473 training, 1,863 validation, and 3,191 testing images. Similar to AnaXNet \cite{anaxnet_miccai2021}, we focus on nine specific findings: lung opacity (LO), pleural effusion (PE), atelectasis (AT), enlarged cardiac silhouette (ECS), pulmonary edema/hazy opacity (PE/HO), pneumothorax (PTX), consolidation (CONS), fluid overload/heart failure (FO/HF) and pneumonia (PN). We consider a radiograph positive for a finding if any region is labelled positive for that finding. We evaluate the retrieval performance by querying each test radiograph with findings against the remaining test set as the database. 

\subsection{Experimental Settings}

In our experiments, we resize all chest radiographs to $224 \times 224$. 
The graph transformer consists of 2 layers with 8 multi-head attention heads and the model dimension is set to $D_T = 64$. 
We utilize the Adam optimizer with a learning rate of $1e-4$ in PyTorch Lightning. We incorporate early stopping based on validation loss with patience of $4$ evaluations. Training is limited to a maximum of $30$ epochs with a batch size of $16$ and gradient accumulation occurs every $8$ epochs. Further, we use a learning rate schedule with a reduction factor of $0.1$ and set gradient clipping to $0.5$. The training process was distributed across four NVIDIA GeForce RTX-$2080$ Ti GPUs. 
\\

\noindent
\textbf{Metrics.} We use three metrics \cite{ath_media2021} to assess the precision and quality of retrieval: (i) Average Precision (AP) measures the average position of relevant cases within the retrieved list, thus evaluating the ranking quality, (ii) Hit Ratio (HR) indicates the proportion of retrieved cases that are relevant, thus reflecting the retrieval effectiveness, (iii) Reciprocal Rank (RR) is the reciprocal of the rank of the first relevant case in the returned list, assessing the retrieval efficiency.

%% file: sections/04_discussion.tex
\section{Results and Discussion}
Various variants of the proposed solution, namely, CheXtriev, were constructed to assess the contributions of various components.  First, we present the evaluation results of these variants. Then, we highlight the strengths of CheXtriev, such as learning from anatomy-aware features via a local approach using the results of benchmarking experiments.\\

\input{tables/variants}

\noindent
\textit{CheXtriev variants.}  The performance results of the variants are presented in Table \ref{variants}. The baseline variant (V0) solely relies on the mean pooling of extracted region features and achieves $51.8\%$ mAP, $39.7\%$ mHR, and $54.0\%$ mRR. The consistent performance boost observed from V1 to V6 (in global MLF), ranges from $+2.1\%$ (in mAP for V1) to $+6.4\%$ (for V6); this underscores the advantage of graph transformers with fully connected unique learnable edges over uniform edge-sharing schemes and naive handcrafted adjacency. This result also suggests that the retrieval task benefits from considering all region pairs, potentially capturing intricate latent relationships between regions. It can be seen that local gated residual connections (V4) lead to a significant drop in performance ($6.38\%$ lower mAP and $6.89\%$ lower mRR) relative to a global one (V6). This emphasizes the value of global gated residual connections with selective refinement for learning multi-level features. The fact that V6 and V3 outperform V5 and V2, respectively, suggests that the learnable location embedding improves model performance by capturing crucial spatial context for accurate ranking.\\

\noindent
\textit{Assessment of anatomy-aware feature extraction.} A trend that can be observed from Table \ref{variants} is that relative to the baseline V0  which lacks inter-region relationship modelling, all graph transformer (GT) variants have a modest but consistent improvement in identifying relevant cases (similar mHR). This suggests that, in general, anatomy-aware feature extraction is more effective (than a strategy that does not consider anatomical relationships) at retrieving relevant cases, and the design of CheXtriev (V6) gives it an additional edge to precisely rank and retrieve these cases (higher mAP and mRR).
\\

\input{tables/benchmarks}

\begin{figure}[t]
    \centering
    \includegraphics[width=0.91\textwidth]{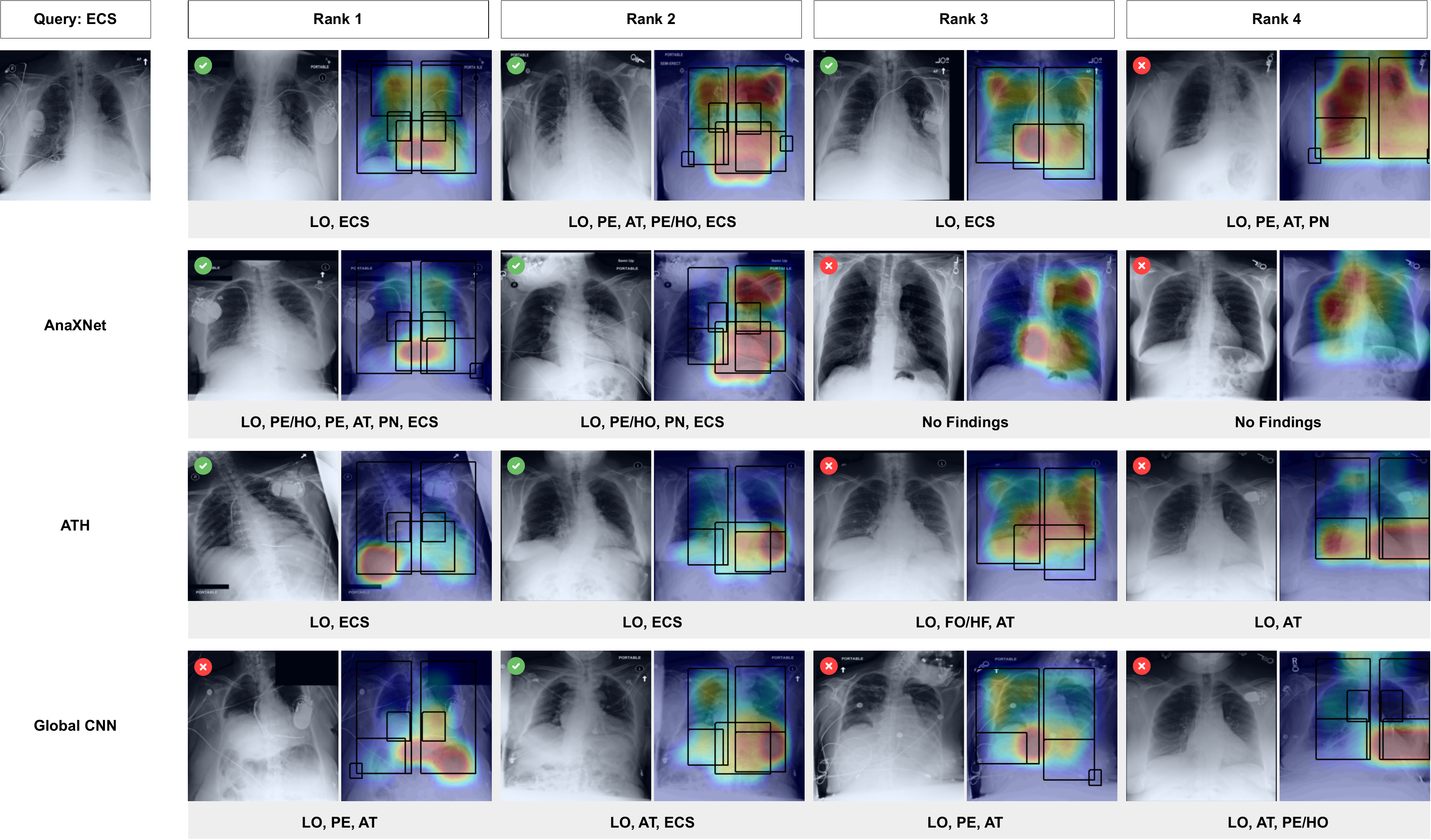}
    \caption{Retrieval performance and saliency map analysis for a sample query image with enlarged cardiac silhouette (ECS). Each row displays the top 4 retrieved images and their corresponding occlusion-based saliency maps generated by CheXtriev, AnaXNet, ATH and Global CNN (top to bottom). A retrieved image is considered correct if it matches the specific finding of interest, in this case ECS, as the query image.} 
    \label{fig:explanations}
\end{figure}

\noindent
\textit{Assessment of global vs. local approaches.} We compare the performance results of CheXtriev, against both global and local methods. Table \ref{benchmarks} presents these results for top five retrieved images. Two global baselines are considered, one based on  ResNet-50 extracted dense features (column 1) and ATH \cite{ath_media2021}, a SOTA retrieval approach (column 2). A student's t-test was done to establish the statistical significance of the difference in results between CheXtriev and the baselines, and the significant values $(p<0.05)$ are marked with an asterisk.  Relative to both the global baselines, CheXtriev has higher AP values across all nine investigated findings, with these ranging from 91.7\% (LO) to 28.8\% (FO/HF). This trend also holds for HR and RR metrics. 
Both macro-mean and weighted mean metrics are reported in the last two rows of Table \ref{benchmarks} to draw insights into the overall retrieval efficacy from an unbalanced database. Macro-mean assigns equal weightage to all findings, which may introduce a bias to frequently occurring classes. Hence, the weighted mean assigns weights to classes inversely proportional to their frequency. It can be observed that the weighted mean is higher than the macro mean for all metrics. 
The mAP improvement over the baselines ranges from at least 12\% (first column) to a notable 23\% (second column). The noteworthy point is that the improvement is quite significant for classes with lower prevalence, such as FO/HF (+82.3\% to +94.6\%), PTX (+40.6\% to +253.9\%), CONS (+30.6\% to +35.6\%), PN (+14.6\% to +20.5\%), demonstrating CheXtriev's ability to learn more discriminative and powerful visual representations compared to global methods.
These results point to CheXtriev's ability to accurately rank and efficiently retrieve relevant cases, even for less frequent classes.\\

\noindent
We also compare CheXtriev against AnaXNet, a SOTA local approach designed for classification tasks. CheXtriev exhibited significant improvements in mean AP, particularly for findings associated with known \textit{blind spots} \cite{missed_diagrad2015,blindspots_sr2015} in chest radiographs, such as lung apices, hilar structures and inferior lung bases (for example, FO/HF +61.80\%, PE/HO +16.82\%, CONS +14.49\% higher). This can be attributed to CheXtriev's global edge-aware attention mechanism and fully connected learnable continuous edges, which enable it to capture anatomical relationships and latent finding co-occurrences more effectively than AnaXNet's naive handcrafted binary edges and limited expressiveness from neighborhood aggregation.
Additionally, CheXtriev's location embeddings and gated residual connections have aided in learning superior visual representations, building upon the strengths of local approaches to capture well the subtle abnormalities in areas prone to human error during interpretation. 
The results also indicate classification-optimized features may not be the most effective for retrieval.
\\

\noindent
\textit{Interpretability analysis.} We analyze saliency maps \cite{xmir_wacv2022} to assess which anatomical regions influenced retrieval for each query image (see Figure \ref{fig:explanations}). The first three retrieved images by CheXtriev all exhibit ECS, with the saliency maps highlighting a focus on the cardiac silhouette region. However, in the saliency map for the fourth retrieved image, the model's attention diffused throughout the lung region, indicating an incorrect retrieval. Some images retrieved by AnaXNet lack any findings for ECS. Even though ATH and CNN manage to retrieve correct cases, the corresponding saliency maps highlight irrelevant regions. This suggests other models might be overly sensitive to visual changes in irrelevant parts of the image, potentially due to weaknesses in encoding spatial information and context within the image. We also note that these saliency maps may not be meaningful for methods that do not use anatomy-aware features, underscoring the additional benefits of our approach.

%% file: tables/variants.tex
\begin{table}[t]
\caption{Impact of design choices in CheXtriev on retrieval and ranking performance. Here, IRM and MLF denote inter-anatomic region modelling using graph transformers and  multi-level features with gated residuals, respectively.}
\fontsize{9}{9}\selectfont
\centering
\begin{tabular}{>{\hspace{0pt}}m{0.12\linewidth}|>{\hspace{0pt}}m{0.1\linewidth}>{\hspace{0pt}}m{0.23\linewidth}>{\hspace{0pt}}m{0.14\linewidth}>{\hspace{0pt}}m{0.10\linewidth}|>{\centering\arraybackslash}m{0.08\linewidth}>{\centering\arraybackslash}m{0.08\linewidth}>{\centering\arraybackslash}m{0.08\linewidth}}
\toprule
Variants & ~IRM & Edge Connectivity & Location & MLF & mAP & mHR & mRR \\ 
\cmidrule(){1-1}\cmidrule(){2-5}\cmidrule(){6-8}
V0 & ~~\xxmark & ~~~~~~~~~\xxmark & ~~~~~\xxmark & ~~~\xxmark & 51.8 & 39.7 & 54.0\\
V1 & ~GT & Shared Binary & ~~~~~\xxmark & Global& 52.9 & 39.9 & 55.2 \\
V2 & ~GT & Shared Uniform & ~~~~~\xxmark & Global& 53.9 & 40.8 & 56.3 \\
V3 & ~GT & Shared Uniform & Learnable & Global& 54.0 & \textbf{40.9} & 56.4 \\
V4 & ~GT & Unique $R_i-R_j$ & Learnable & Local& 51.8 & 39.8 & 53.7 \\
V5 & ~GT & Unique $R_i-R_j$ & ~~~~~\xxmark & Global &  54.6& \textbf{40.9} & 57.1 \\
V6 & ~GT & Unique $R_i-R_j$ & Learnable & Global & \textbf{55.1} & 40.7 & \textbf{57.4} \\
\bottomrule
\end{tabular}
\label{variants}
\end{table}

%% file: tables/benchmarks.tex
\begin{table*}[t]
\caption{Comparison of top-5 retrieval performance on MIMIC-CXR dataset against global baselines (CNN, ATH) and a local variant (AnaXNet). $(.)^{\ast}$ indicates $p < 0.05$.}
\centering
\fontsize{8.5}{9}\selectfont
\begin{tabular}{
>{\arraybackslash}p{0.13\linewidth}
*{12}{>{\centering\arraybackslash}p{0.064\linewidth}}
} 
\toprule
      & 
      \multicolumn{3}{>{\centering}m{0.201\linewidth}}{Global CNN} & \multicolumn{3}{>{\centering}m{0.201\linewidth}}{ATH \cite{ath_media2021}} 
      & \multicolumn{3}{>{\centering}m{0.201\linewidth}}{AnaXNet \cite{anaxnet_miccai2021}} 
      & \multicolumn{3}{>{\centering}m{0.201\linewidth}}{CheXtriev}  \\ 
\cmidrule(r){1-1}\cmidrule(lr){2-4}\cmidrule(lr){5-7}\cmidrule(lr){8-10}\cmidrule(lr){11-13}
Findings & AP & HR & RR & AP & HR & RR & AP & HR & RR & AP & HR & RR \\
\midrule
LO   & $90.4^{\tiny\ast}$ & $85.9^{\tiny\ast}$ & 92.2 & $88.2^{\tiny\ast}$ & $82.2^{\tiny\ast}$ & 90.9 & $88.8^{\tiny\ast}$ & $82.9^{\tiny\ast}$ & 91.2 & \textbf{91.7} & \textbf{87.6} & \textbf{93.3} \\
PE   & $68.6^{\tiny\ast}$ & $54.7^{\tiny\ast}$ & 72.6 & $62.1^{\tiny\ast}$ & $46.3^{\tiny\ast}$ & 66.1 & $63.7^{\tiny\ast}$ & $46.2^{\tiny\ast}$ & 67.2 & \textbf{71.4} & \textbf{59.4} & \textbf{74.7} \\
AT   & $60.3^{\tiny\ast}$ & $44.1^{\tiny\ast}$ & 64.2 & $54.3^{\tiny\ast}$ & $36.8^{\tiny\ast}$ & 57.8 & $59.1^{\tiny\ast}$ & $42.3^{\tiny\ast}$ & 62.8 & \textbf{64.4} & \textbf{49.1} & \textbf{67.6} \\
ECS  & $64.7^{\tiny\ast}$ & $48.3^{\tiny\ast}$ & 67.9 & $61.3^{\tiny\ast}$ & $45.4^{\tiny\ast}$ & 64.9 & $64.7^{\tiny\ast}$ & $48.5^{\tiny\ast}$ & 68.4 & \textbf{69.4} & \textbf{56.0} & \textbf{73.7} \\
PE/HO & $56.5^{\tiny\ast}$ & $40.2^{\tiny\ast}$ & 59.2 & $51.8^{\tiny\ast}$ & $34.7^{\tiny\ast}$ & 54.7 & $53.5^{\tiny\ast}$ & $34.9^{\tiny\ast}$ & 57.4 & \textbf{62.5} & \textbf{47.8} & \textbf{65.8} \\
PTX  & $22.4^{\tiny\ast}$ & $8.5^{\tiny\ast}$ & 22.7 & $8.9^{\tiny\ast}$ & $4.0^{\tiny\ast}$ & 9.1 & 31.1 & \textbf{12.7} & 31.8 & \textbf{31.5} & 12.5 & \textbf{32.4} \\
CONS & $24.2^{\tiny\ast}$ & 13.2 & 25.2 & $23.3^{\tiny\ast}$ & $11.8^{\tiny\ast}$ & 23.8 & 27.6 & 12.5 & 28.3 & \textbf{31.6} & \textbf{14.6} & \textbf{32.7} \\
FO/HF & $14.8^{\tiny\ast}$ & $7.3^{\tiny\ast}$ & 14.8 & $15.8^{\tiny\ast}$ & $7.0^{\tiny\ast}$ & 16.2 & $17.8^{\tiny\ast}$ & $7.2^{\tiny\ast}$ & 17.9 & \textbf{28.8} & \textbf{11.8} & \textbf{29.3} \\
PN   & $39.0^{\tiny\ast}$ & $23.0^{\tiny\ast}$ & 41.3 & $37.1^{\tiny\ast}$ & $21.9^{\tiny\ast}$ & 38.8 & $41.4^{\tiny\ast}$ & $23.6^{\tiny\ast}$ & 43.4 & \textbf{44.7} & \textbf{27.1} & \textbf{47.2} \\
\midrule
Mean   & 49.0 & 36.1 & 51.1 & 44.8 & 32.2 & 46.9 & 49.7 & 34.5 & 52.0 & \textbf{55.1} & \textbf{40.7} & \textbf{57.4} \\
wMean  & 67.0 & 55.2 & 69.6 & 63.1 & 50.5 & 66.1 & 65.7 & 52.2 & 68.6 & \textbf{70.8} & \textbf{59.5} & \textbf{73.4} \\
\bottomrule
\end{tabular}
\label{benchmarks}
\end{table*}

%% file: sections/05_conclusion.tex
\section{Conclusion}

In this work, we propose CheXtriev, a graph-based radiograph retrieval model inspired by the systematic approach radiologists use to interpret radiographs and grounded in evidence-based anatomy. Key novel features of our approach were explicitly targeting informative details from specific anatomical regions, modelling the interplay between anatomical location and findings and fusing them into a richer anatomy-aware representation. Superior results over existing methods underscore the benefit of this contextualization in achieving more accurate, effective, and efficient case retrieval, particularly for the less prevalent findings. A preliminary analysis with anatomy-aware saliency maps indicates that it may be possible to use them,  albeit with some further extension to cover multiple findings, for interpretability of the retrieved results.
Overall, CheXtriev offers a promising approach for medical image retrieval tasks, particularly in chest radiography, where subtle anatomical variations hold significant diagnostic value.

\begin{credits}
\subsubsection{\ackname} We thank Dr. L.T. Kishore for his insights on how radiologists analyze and interpret chest radiographs in clinical practice.
\subsubsection{\discintname}
There are no conflicts of interest to declare. 
\end{credits}

%% file: sections/06_bibliography.tex
%
%
%
\bibliographystyle{splncs04}
%

%% file: SupplementaryFile-3636/SupplementaryLatexSource-3636/supplementarypaper.tex
\title{CheXtriev: Anatomy-Centered Representation for Case-Based Retrieval of Chest Radiographs}
\blfootnote{N. Akash and A. Tadanki contributed equally.}
\subtitle{Supplementary Material}

\titlerunning{CheXtriev: Anatomy-Centered Case-Based Retrieval of CXRs}

\author{Naren Akash R J \index{Akash, Naren}\orcidlink{0000-0002-5218-8434}\and
Arihanth Tadanki \index{Tadanki, Arihanth}\orcidlink{0009-0006-5554-1703} \and
Jayanthi Sivaswamy \index{Sivaswamy, Jayanthi}\orcidlink{0000-0001-9588-0482}
}

\authorrunning{N. Akash et al.}

\institute{International Institute of Information Technology Hyderabad, India}

\maketitle

\input{SupplementaryFile-3636/SupplementaryLatexSource-3636/top3}
\input{SupplementaryFile-3636/SupplementaryLatexSource-3636/top10}
\input{SupplementaryFile-3636/SupplementaryLatexSource-3636/testdb}
\input{SupplementaryFile-3636/SupplementaryLatexSource-3636/bibliography}

%% file: SupplementaryFile-3636/SupplementaryLatexSource-3636/top3.tex
\begin{table*}[h]
\caption{Comparison of \textbf{top-3} retrieval performance on MIMIC-CXR dataset against global baselines (CNN, ATH) and a local variant (AnaXNet). $(.)^{\textcolor{blue}{\ast}}$ indicates $p < 0.05$.}
\centering
\fontsize{8.5}{9}\selectfont
\begin{tabular}{
>{\arraybackslash}p{0.13\linewidth}
*{12}{>{\centering\arraybackslash}p{0.064\linewidth}}
} 
\toprule
      & 
      \multicolumn{3}{>{\centering}m{0.201\linewidth}}{Global CNN} & \multicolumn{3}{>{\centering}m{0.201\linewidth}}{ATH \cite{ath_media2021}} 
      & \multicolumn{3}{>{\centering}m{0.201\linewidth}}{AnaXNet \cite{anaxnet_miccai2021}} 
      & \multicolumn{3}{>{\centering}m{0.201\linewidth}}{CheXtriev}  \\ 

\cmidrule(r){1-1}\cmidrule(lr){2-4}\cmidrule(lr){5-7}\cmidrule(lr){8-10}\cmidrule(lr){11-13}
Findings & AP & HR & RR & AP & HR & RR & AP & HR & RR & AP & HR & RR \\
\midrule
LO   & 
91.4$^{\textcolor{blue}{\ast}}$ & 
85.8$^{\textcolor{blue}{\ast}}$ & 
91.9$^{\textcolor{blue}{\ast}}$ & 
89.7$^{\textcolor{blue}{\ast}}$ & 
82.5$^{\textcolor{blue}{\ast}}$ & 
90.6$^{\textcolor{blue}{\ast}}$ & 
90.7$^{\textcolor{blue}{\ast}}$ & 
84.0$^{\textcolor{blue}{\ast}}$ & 
91.7$^{\textcolor{blue}{\ast}}$ & 
\textbf{92.5} & 
\textbf{88.2} & 
\textbf{93.0} \\

PE   & 
70.0$^{\textcolor{blue}{\ast}}$ & 
56.2$^{\textcolor{blue}{\ast}}$ & 
71.2 & 
62.6$^{\textcolor{blue}{\ast}}$ & 
47.6$^{\textcolor{blue}{\ast}}$ & 
63.9$^{\textcolor{blue}{\ast}}$ & 
64.8$^{\textcolor{blue}{\ast}}$ & 
48.6$^{\textcolor{blue}{\ast}}$ & 
65.7$^{\textcolor{blue}{\ast}}$ & 
\textbf{72.1} & 
\textbf{59.4} & 
\textbf{73.0} \\

AT   & 
60.8$^{\textcolor{blue}{\ast}}$ & 
45.2$^{\textcolor{blue}{\ast}}$ & 
61.9$^{\textcolor{blue}{\ast}}$ & 
53.9$^{\textcolor{blue}{\ast}}$ & 
36.8$^{\textcolor{blue}{\ast}}$ & 
54.5$^{\textcolor{blue}{\ast}}$ & 
59.7$^{\textcolor{blue}{\ast}}$ & 
42.5$^{\textcolor{blue}{\ast}}$ & 
61.1$^{\textcolor{blue}{\ast}}$ & 
\textbf{64.9} & 
\textbf{48.5} & 
\textbf{65.8} \\

ECS  & 
64.5$^{\textcolor{blue}{\ast}}$ & 
49.3$^{\textcolor{blue}{\ast}}$ & 
65.6$^{\textcolor{blue}{\ast}}$ & 
61.6$^{\textcolor{blue}{\ast}}$ & 
45.9$^{\textcolor{blue}{\ast}}$ & 
62.5$^{\textcolor{blue}{\ast}}$ & 
64.5$^{\textcolor{blue}{\ast}}$ & 
48.3$^{\textcolor{blue}{\ast}}$ & 
65.2$^{\textcolor{blue}{\ast}}$ & 
\textbf{70.5} & 
\textbf{55.4} & 
\textbf{71.7} \\

PE/HO & 
55.3$^{\textcolor{blue}{\ast}}$ & 
40.7$^{\textcolor{blue}{\ast}}$ & 
56.1$^{\textcolor{blue}{\ast}}$ & 
51.0$^{\textcolor{blue}{\ast}}$ & 
36.0$^{\textcolor{blue}{\ast}}$ & 
51.7$^{\textcolor{blue}{\ast}}$ & 
53.7$^{\textcolor{blue}{\ast}}$ & 
36.3$^{\textcolor{blue}{\ast}}$ & 
54.5$^{\textcolor{blue}{\ast}}$ & 
\textbf{64.0} & 
\textbf{49.1} & 
\textbf{65.2} \\

PTX  & 
20.7$^{\textcolor{blue}{\ast}}$ & 
11.5$^{\textcolor{blue}{\ast}}$ & 
21.0$^{\textcolor{blue}{\ast}}$ & 
7.6$^{\textcolor{blue}{\ast}}$ & 
4.2$^{\textcolor{blue}{\ast}}$ & 
7.8$^{\textcolor{blue}{\ast}}$ & 
32.0 & 
\textbf{16.3} & 
32.1 & 
\textbf{29.7} & 
14.9 & 
\textbf{29.7} \\

CONS & 
21.6$^{\textcolor{blue}{\ast}}$ & 
12.9 & 
21.8$^{\textcolor{blue}{\ast}}$ & 
20.9$^{\textcolor{blue}{\ast}}$ & 
13.0$^{\textcolor{blue}{\ast}}$ & 
21.1$^{\textcolor{blue}{\ast}}$ & 
25.6 & 
15.0 & 
25.8 & 
\textbf{27.2} & 
\textbf{15.9} & 
\textbf{27.7} \\

FO/HF & 
11.2$^{\textcolor{blue}{\ast}}$ & 
6.5$^{\textcolor{blue}{\ast}}$ & 
11.3$^{\textcolor{blue}{\ast}}$ & 
13.6$^{\textcolor{blue}{\ast}}$ & 
7.3$^{\textcolor{blue}{\ast}}$ & 
13.7$^{\textcolor{blue}{\ast}}$ & 
15.0$^{\textcolor{blue}{\ast}}$ & 
7.7$^{\textcolor{blue}{\ast}}$ & 
15.1$^{\textcolor{blue}{\ast}}$ & 
\textbf{26.3} & 
\textbf{13.6} & 
\textbf{26.4} \\

PN   & 
37.1$^{\textcolor{blue}{\ast}}$ & 
22.8$^{\textcolor{blue}{\ast}}$ & 
37.5$^{\textcolor{blue}{\ast}}$ & 
34.3$^{\textcolor{blue}{\ast}}$ & 
21.4$^{\textcolor{blue}{\ast}}$ & 
34.6$^{\textcolor{blue}{\ast}}$ & 
39.4$^{\textcolor{blue}{\ast}}$ & 
26.2$^{\textcolor{blue}{\ast}}$ & 
39.7$^{\textcolor{blue}{\ast}}$ & 
\textbf{44.8} & 
\textbf{28.2} & 
\textbf{45.4} \\

\midrule
Mean   & 48.1 & 36.8 & 48.7 & 43.9 & 32.7 & 44.5 & 49.5 & 36.1 & 50.1 & \textbf{54.7} & \textbf{41.5} & \textbf{55.3} \\
wMean  & 67.1 & 55.7 & 67.9 & 63.2 & 51.0 & 64.0 & 66.3 & 53.4 & 67.1 & \textbf{71.3} & \textbf{59.8} & \textbf{72.0} \\

\bottomrule
\end{tabular}
\label{benchmarks}
\end{table*}

%% file: SupplementaryFile-3636/SupplementaryLatexSource-3636/top10.tex
\begin{table*}[h]
\caption{Comparison of \textbf{top-10} retrieval performance on MIMIC-CXR dataset against global baselines (CNN, ATH) and a local variant (AnaXNet). $(.)^{\textcolor{blue}{\ast}}$ indicates $p < 0.05$.}
\centering
\fontsize{8.5}{9}\selectfont
\begin{tabular}{
>{\arraybackslash}p{0.13\linewidth}
*{12}{>{\centering\arraybackslash}p{0.064\linewidth}}
} 
\toprule
      & 
      \multicolumn{3}{>{\centering}m{0.201\linewidth}}{Global CNN} & \multicolumn{3}{>{\centering}m{0.201\linewidth}}{ATH \cite{ath_media2021}} 
      & \multicolumn{3}{>{\centering}m{0.201\linewidth}}{AnaXNet \cite{anaxnet_miccai2021}} 
      & \multicolumn{3}{>{\centering}m{0.201\linewidth}}{CheXtriev}  \\ 

\cmidrule(r){1-1}\cmidrule(lr){2-4}\cmidrule(lr){5-7}\cmidrule(lr){8-10}\cmidrule(lr){11-13}
Findings & AP & HR & RR & AP & HR & RR & AP & HR & RR & AP & HR & RR \\
\midrule
LO   & 
89.0$^{\textcolor{blue}{\ast}}$ & 
86.1$^{\textcolor{blue}{\ast}}$ & 
92.2$^{\textcolor{blue}{\ast}}$ & 
86.2$^{\textcolor{blue}{\ast}}$ & 
82.0$^{\textcolor{blue}{\ast}}$ & 
90.9$^{\textcolor{blue}{\ast}}$ & 
87.3$^{\textcolor{blue}{\ast}}$ & 
83.5$^{\textcolor{blue}{\ast}}$ & 
91.9$^{\textcolor{blue}{\ast}}$ & 
\textbf{90.5} & 
\textbf{87.7} & 
\textbf{93.2} \\

PE   & 
64.4$^{\textcolor{blue}{\ast}}$ & 
54.1$^{\textcolor{blue}{\ast}}$ & 
73.2 & 
58.2$^{\textcolor{blue}{\ast}}$ & 
45.2$^{\textcolor{blue}{\ast}}$ & 
67.2$^{\textcolor{blue}{\ast}}$ & 
59.3$^{\textcolor{blue}{\ast}}$ & 
46.4$^{\textcolor{blue}{\ast}}$ & 
68.9$^{\textcolor{blue}{\ast}}$ & 
\textbf{67.6} & 
\textbf{57.5} & 
\textbf{75.1} \\

AT   & 
55.7$^{\textcolor{blue}{\ast}}$ & 
43.5$^{\textcolor{blue}{\ast}}$ & 
65.0$^{\textcolor{blue}{\ast}}$ & 
50.5$^{\textcolor{blue}{\ast}}$ & 
36.1$^{\textcolor{blue}{\ast}}$ & 
59.2$^{\textcolor{blue}{\ast}}$ & 
55.1$^{\textcolor{blue}{\ast}}$ & 
41.4$^{\textcolor{blue}{\ast}}$ & 
65.0$^{\textcolor{blue}{\ast}}$ & 
\textbf{59.7} & 
\textbf{47.6} & 
\textbf{68.6} \\

ECS  & 
59.8$^{\textcolor{blue}{\ast}}$ & 
48.5$^{\textcolor{blue}{\ast}}$ & 
68.5$^{\textcolor{blue}{\ast}}$ & 
57.3$^{\textcolor{blue}{\ast}}$ & 
43.8$^{\textcolor{blue}{\ast}}$ & 
65.7$^{\textcolor{blue}{\ast}}$ & 
59.5$^{\textcolor{blue}{\ast}}$ & 
46.1$^{\textcolor{blue}{\ast}}$ & 
68.3$^{\textcolor{blue}{\ast}}$ & 
\textbf{64.6} & 
\textbf{53.1} & 
\textbf{74.1} \\

PE/HO & 
52.3$^{\textcolor{blue}{\ast}}$ & 
39.1$^{\textcolor{blue}{\ast}}$ & 
60.2$^{\textcolor{blue}{\ast}}$ & 
48.4$^{\textcolor{blue}{\ast}}$ & 
32.8$^{\textcolor{blue}{\ast}}$ & 
56.1$^{\textcolor{blue}{\ast}}$ & 
50.2$^{\textcolor{blue}{\ast}}$ & 
34.0$^{\textcolor{blue}{\ast}}$ & 
59.1$^{\textcolor{blue}{\ast}}$ & 
\textbf{59.2} & 
\textbf{46.3} & 
\textbf{68.0} \\

PTX  & 
23.6 & 
6.8$^{\textcolor{blue}{\ast}}$ & 
24.6$^{\textcolor{blue}{\ast}}$ & 
10.3$^{\textcolor{blue}{\ast}}$ & 
4.3$^{\textcolor{blue}{\ast}}$ & 
11.3$^{\textcolor{blue}{\ast}}$ & 
\textbf{32.1} & 
9.5$^{\textcolor{blue}{\ast}}$ & 
35.3 & 
31.4 & 
\textbf{11.5} & 
\textbf{35.4} \\

CONS & 
26.3$^{\textcolor{blue}{\ast}}$ & 
13.2 & 
28.9 & 
24.8$^{\textcolor{blue}{\ast}}$ & 
10.8$^{\textcolor{blue}{\ast}}$ & 
26.2$^{\textcolor{blue}{\ast}}$ & 
27.2 & 
11.3 & 
30.7 & 
\textbf{30.7} & 
\textbf{14.1} & 
\textbf{33.8} \\

FO/HF & 
16.5$^{\textcolor{blue}{\ast}}$ & 
6.7$^{\textcolor{blue}{\ast}}$ & 
16.9$^{\textcolor{blue}{\ast}}$ & 
16.6$^{\textcolor{blue}{\ast}}$ & 
6.7$^{\textcolor{blue}{\ast}}$ & 
18.3$^{\textcolor{blue}{\ast}}$ & 
18.9$^{\textcolor{blue}{\ast}}$ & 
6.3$^{\textcolor{blue}{\ast}}$ & 
20.3$^{\textcolor{blue}{\ast}}$ & 
\textbf{29.8} & 
\textbf{10.7} & 
\textbf{32.1} \\

PN   & 
38.1$^{\textcolor{blue}{\ast}}$ & 
22.3$^{\textcolor{blue}{\ast}}$ & 
44.2$^{\textcolor{blue}{\ast}}$ & 
36.2$^{\textcolor{blue}{\ast}}$ & 
22.2$^{\textcolor{blue}{\ast}}$ & 
41.6$^{\textcolor{blue}{\ast}}$ & 
39.5$^{\textcolor{blue}{\ast}}$ & 
22.4$^{\textcolor{blue}{\ast}}$ & 
45.5$^{\textcolor{blue}{\ast}}$ & 
\textbf{42.8} & 
\textbf{22.5} & 
\textbf{50.7} \\

\midrule
Mean   & 47.3 & 35.6 & 52.6 & 43.2 & 31.5 & 48.5 & 47.7 & 33.4 & 53.9 & \textbf{52.9} & \textbf{39.3} & \textbf{59.0} \\
wMean  & 64.2 & 54.9 & 70.5 & 60.5 & 49.7 & 67.1 & 62.8 & 51.7 & 69.9 & \textbf{67.9} & \textbf{58.3} & \textbf{74.3} \\

\bottomrule
\end{tabular}
\label{benchmarks_top10}
\end{table*}

%% file: SupplementaryFile-3636/SupplementaryLatexSource-3636/testdb.tex
\begin{table}[t]
\caption{Statistics of the Chest ImageGenome \cite{chestimagenome_neurips2021} test dataset, which includes frontal chest radiographs (PA or AP view) with valid bounding box annotations for 18 anatomical regions. The data split is based on the official MIMIC-CXR \cite{mimiccxr_scidata2019,mimiccxrjpg_2019} data splits.}
\fontsize{9}{9}\selectfont
\centering
\begin{tabular}{l|r}
\toprule
\quad Radiological Findings &  \quad \#Radiographs \quad  \\
\midrule
\quad Lung Opacity       \quad   &     2679  \quad   \\
\quad Pleural Effusion     \quad   &     1340    \quad  \\
\quad Atelectasis      \quad     &    1241   \quad  \\
\quad Enlarged Cardiac Silhouette  \quad &     1212     \quad\\
\quad Pulmonary Edema / Hazy Opacity  \quad &     819  \quad  \\
\quad Pneumonia       \quad     &     643    \quad \\
\quad Consolidation        \quad  &     334      \quad\\
\quad Fluid Overload / Heart Failure  \quad &     169    \quad  \\
\quad Pneumothorax      \quad    &      96    \quad \\
\bottomrule
\end{tabular}
\label{testdb}
\end{table}

%% file: SupplementaryFile-3636/SupplementaryLatexSource-3636/bibliography.tex
\bibliographystyle{splncs04}